\documentclass[reprint,aps,nofootinbib,superscriptaddress,preprintnumbers]{revtex4-2}

\usepackage{braket}
\usepackage{slashed}
\usepackage{graphicx}
\usepackage{xcolor}
\usepackage{enumitem}
\usepackage{siunitx}
\usepackage{mathtools}
\usepackage{dsfont}
\usepackage[hidelinks]{hyperref}
\usepackage{amsmath,amsthm,amsfonts,amscd,amssymb}

\newcommand\blfootnote[1]{%
  \begingroup
  \renewcommand\thefootnote{}\footnote{#1}%
  \addtocounter{footnote}{-1}%
  \endgroup
}

\usepackage{tikz} 
\usetikzlibrary{shapes,arrows,positioning,automata,backgrounds,calc,er,patterns}
\usetikzlibrary{decorations,decorations.markings,decorations.pathreplacing,decorations.pathmorphing}
\usepackage{relsize}

\newcommand{\be}{\begin{equation}}
\newcommand{\ee}{\end{equation}} 
\newcommand{\mb}{\mathbf}

\usepackage{tensor}

\begin{document}

\preprint{CALT-TH 2024-034}

\title{Response of interferometers to the vacuum of quantum gravity}

\author{Daniel Carney}
\affiliation{Physics Division, Lawrence Berkeley National Laboratory, Berkeley, CA}
\author{Manthos Karydas}
\affiliation{Physics Division, Lawrence Berkeley National Laboratory, Berkeley, CA}
\author{Allic Sivaramakrishnan}
\affiliation{Walter Burke Institute for Theoretical Physics, California Institute of Technology, Pasadena, CA}

\blfootnote{carney@lbl.gov, mkarydas@lbl.gov, allic@caltech.edu \\}

\begin{abstract}
    
It has recently been suggested that exotic quantum gravity effects could lead to large vacuum fluctuations, potentially observable with realistic detectors. Experiments are currently being built to search for these signals. Here we analyze the minimal model of quantum gravity at low energies --- the usual effective quantum field theory of gravitons --- and show that it unambiguously predicts an unobservably small variation in the measured interferometer length $\Delta L \sim \ell_{\rm pl} \sim 10^{-35}~{\rm m}$. In particular, there are no divergences signaling a breakdown of this calculation in the low energy regime. Thus, detection of a large, gravitationally-induced length variation would signal a severe breakdown of effective quantum field theory in low energy quantum gravity.

\end{abstract}

\maketitle

Assuming the gravitational field $g_{\mu\nu}$ is a quantum mechanical degree of freedom, it should exhibit fluctuations, even in its vacuum state. These fluctuations could in principle be measured by a gravitational wave detector such as an interferometer, manifesting as random variations $\Delta L$ around the equilibrium length $L$ of the device. At present, Advanced LIGO \cite{aasi2015advanced} is the most sensitive such device available, and is capable of detecting $h \sim \Delta L/L \approx 10^{-23}$ in the 10 Hz---10 kHz band, where $h$ measures small deviations around flat spacetime $g_{\mu\nu} = \eta_{\mu\nu} + h_{\mu\nu}$. Many new interferometers are currently being constructed with similar sensitivity up to MHz scales \cite{Aggarwal:2020olq,Aggarwal:2020umq,vermeulen2021experiment,Vermeulen:2024vgl}.

What is the expected scale of $\Delta L/L$ due to quantum fluctuations in the spacetime metric? Somewhat controversially, it has periodically been suggested \cite{Hogan:2007hc,hogan2008measurement,Verlinde:2019xfb,Verlinde:2019ade,Zurek:2020ukz,Li:2022mvy,Bub:2023bfi} that exotic effects in quantum gravity could lead to random fluctuations of order $\Delta L/L \sim \sqrt{\ell_{\rm pl}/L} \sim 10^{-19}$ (for the LIGO baseline $L = 4~{\rm km}$), well within the regime of observability. Indeed, such a model is only consistent with observed LIGO data if the fluctuations are at frequencies outside the LIGO band, as claimed in \cite{Li:2022mvy,Bub:2023bfi}. 

Such large fluctuations would be counter to intuition from basic effective field theory (EFT) arguments. Under the minimal hypothesis that gravitational perturbations $h_{\mu\nu}$ are quantized into gravitons and treated as an EFT~\cite{Donoghue:1995cz,Burgess:2003jk,Donoghue:2022eay}, one would expect fluctuations of order $\Delta L/L \sim \ell_{\rm pl}/L \sim 10^{-38}$, far too small to be observed~\cite{Jaekel:1993ft,Pang:2018eec,Parikh:2020nrd,Parikh:2020kfh,Parikh:2020fhy,Hertzberg:2021rbl,Guerreiro:2021qgk,Carney:2023nzz}. This theory should be valid for experiments at energy densities below the Planck scale, which is certainly the situation in a terrestrial interferometer. 

Our goal in this paper is to provide a simple and rigorous calculation of the expected length fluctuations in the effective field theory picture. In particular, a question which has previously been left open is whether the EFT calculation may contain divergence~\cite{Parikh:2020nrd,Parikh:2020kfh,Parikh:2020fhy}, either in the ultraviolet or infrared, which signal that the EFT calculation breaks down at low energies and needs to be augmented by some non-perturbative effect which could lead to large fluctuations.\footnote{Many of the suggestions of large fluctuations have been based on heuristic arguments citing ideas from holography. It should be emphasized that the only known controlled holographic models are string theories. These contain a massless spin-2 boson in the low-energy spectrum and therefore must actually perturbatively reproduce the basic EFT prediction $\Delta L/L \sim \ell_{\rm pl}/L$ \cite{weinberg1964photons,weinberg1965photons,Maldacena:1997re,Heemskerk:2009pn}.} 

We begin by defining a gauge-invariant length observable and discuss its vacuum fluctuations. We perform a careful study of any possible uncontrollable divergences which might signal the breakdown of the EFT calculation.  We find no such divergences and therefore conclude that the EFT gives a well-defined, unambiguous prediction at leading order in the graviton expansion. We also give a detailed prediction for the resulting signal strength in realistic detectors, such as LIGO and GQuEST, and show that these fluctuations are unobservably small (see Fig. \ref{fig:psds}). This means that the scaling $\Delta L/L \sim \sqrt{\ell_{\rm pl}/L}$ suggested in \cite{Hogan:2007hc,hogan2008measurement,Verlinde:2019xfb,Verlinde:2019ade,Zurek:2020ukz,Li:2022mvy,Bub:2023bfi}, if observed, would require that the metric is not quantized perturbatively into gravitons according to the usual rules of field theory, even at low energies accessible in laboratory experiments.

\section{Photon time-of-flight and length}

The fundamental observable in an interferometer is the phase of light which traverses a path of unknown length. This is used to infer the length of the path by noting that the light phase $\phi$ changes by an amount proportional to the path length $\phi = 2L/\lambda = \tau \omega_{\ell}/2\pi$, where $\lambda = 2\pi/\omega_{\ell}$ is the wavelength of laser light at frequency $\omega_{\ell}$ and $\tau$ is the time-of-flight of the photon. Thus, ultimately, what we will want to do is make a prediction for the noise in the output light phase in the interferometer, and use this to estimate the uncertainty of the inferred interferometer length. We give a full quantum calculation of this observable in a LIGO-like interferometer in Sec. \ref{sec:detector}.

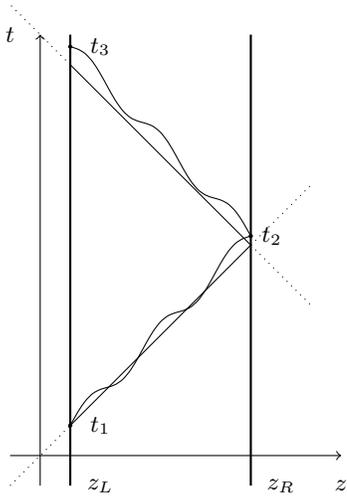
\begin{figure}[t]
\begin{tikzpicture}[scale=0.8]

\draw [->] (-.5,0) -- (5,0);
\draw [->] (0,-.5) -- (0,7);

\node at (5,-.5) {$z$};
\node at (-.5,7) {$t$};

\draw [thick] (.5,-.5) -- (.5,7);
\draw [thick] (3.5,-.5) -- (3.5,7);

\node at (1,-.5) {$z_L$};
\node at (4,-.5) {$z_R$};

\draw (.5,.5) -- (3.5,3.5);
\draw (3.5,3.5) -- (.5,6.5);
\draw [dotted] (3.5,3.5) -- (4.5,4.5);
\draw [dotted] (3.5,3.5) -- (4.5,2.5);
\draw [dotted] (-.5,-.5) -- (.5,.5);
\draw [dotted] (.5,6.5) -- (-.5,7.5);

\pgfmathsetmacro{\epsilon}{0.15}
\draw [domain=0:1, smooth, variable=\t] 
    plot ({0.5 + 3*\t}, {0.5 + (3+ \epsilon)*\t + \epsilon*sin(5*\t*pi r)});
\draw [domain=0:1, smooth, variable=\t] 
    plot ({3.5 - 3*\t}, {3.5+\epsilon + (3+ \epsilon)*\t  + \epsilon*sin(5*\t*pi r)});

\node at (1,.5) {$t_1$};
\node at (4-\epsilon,3.5+\epsilon) {$t_2$};
\node at (1,6.5+2*\epsilon) {$t_3$};

\fill (.5,.5) circle (1pt); 
\fill (3.5,3.5+\epsilon) circle (1pt); 
\fill (.5,6.5+2*\epsilon) circle (1pt); 

\end{tikzpicture}
\caption{Interferometer geometry in spacetime. The oscillating line represents the geodesic path of the photon in the presence of $O(h)$ metric fluctuations. The line can also extend into the transverse 
$x,y$ plane, which is suppressed in this figure. The arrival time $t_{3}$ has an $O(h)$ offset compared to the straight unperturbed path. The proper time $\tau(t_{1})\equiv t_{3}-t_{1}$ is given in Eq. \eqref{eq:round-trip}.}
\label{fig:interferometer}
\end{figure}

To warm up, we can start with a more geometrical calculation. Consider a pair of freely falling mirrors, initially at rest, with light between them, as in Fig. \ref{fig:interferometer}. The proper time $\tau$ elapsed on the worldline of the left mirror between photon emission at $t_1$ and absorption at $t_3$ is (perturbatively) gauge-invariant because it is defined by the intersections of geodesics \cite{Dimopoulos:2007cj,Lee:2024oxo}. For related proper time observables and their correlators, see \cite{Sivaramakrishnan:2024ydy}.

We now compute $\tau$ and its noise properties. We will work in the transverse-traceless gauge, in which gravitational waves do not cause the mirrors to change their coordinate positions $z_{L,R}$ to leading order in $h$ \cite{Maggiore:2007ulw}. For the metric in the TT gauge
\be
\label{metric_TT_gauge}
ds^{2}=- dt^{2}+ (\delta_{ij}+ h_{ij})dx^{i}dx^{j}\,,
\ee
the photon path will receive $O(h)$ corrections to the null geodesic in flat Minkowski spacetime. A simple calculation (see Appendix \ref{app:geodesics} for a few details) shows that to $O(h)$, the time elapsed on an inertial clock at $z_L$ reads
\be
\label{eq:round-trip}
\tau(t_{1})=2L+ \frac{1}{2}\int^{t_{1}+ 2L}_{t_{1}}dt\, h_{zz}(t,z(t)) + O(h^2)\,,
\ee
where $z(t)$ is the piecewise linear geodesic in Minkowski coordinates, and $x=y=0$ in the argument of the metric. In perturbative quantum gravity, $h_{\mu\nu}$ is promoted to an operator through canonical quantization:
\be
\label{eq:hmunu}
h_{\mu\nu}(x) = \ell_{\rm Pl} \sum_{s} \int \frac{d^3\mb{k}}{\sqrt{(2\pi)^3 2 E_{\mb{k} }}} \left[ \epsilon_{\mu\nu}^s(\mb{k}) e^{i k \cdot x} b_{\mb{k},s} + {\rm h.c.} \right],
\ee
where $s = 1,2$ labels polarization states, and the canonical commutation relation is $[ b_{\mb{k},s}, b_{\mb{k}',s'}^\dag ] = \delta^3(\mb{k}-\mb{k}')\delta_{ss'}$. We have normalized this with the Planck length $\ell_{\rm Pl} = \sqrt{32 \pi G_N}$ so that $h$ is dimensionless, i.e., has units of strain. Thus $\tau$ becomes a non-local operator on the gravitational Hilbert space. In the vacuum $b_{\mb{k},s} \ket{0} = 0$, we have $\braket{\tau} = 2L$. 

What we are interested in is the level of noise in $\tau$. The simplest characterization of this would be to compute the variance $\braket{\Delta \tau^2}$, where $\Delta \tau = \tau - \braket{\tau}$. This turns out to be UV divergent for the usual reason that it involves coincident points of a two-point correlation function. To study this, let us instead work out the more general correlator
\begin{align}
\begin{split}
\label{eq:tautau}
\braket{\tau(t)\tau(0)} & = \frac{1}{4} \int \displaylimits_{t}^{t+2L} dt' \int \displaylimits_{0}^{2L} dt'' \braket{h_{zz}(t',z(t')) h_{zz}(t'',z(t'')) } \\
& = \ell_{\rm Pl}^2  \sum_s \int \frac{d^3\mb{k}}{(2\pi)^3 2 E_{\mb{k}}} | \epsilon_{zz}^s (\mb{k})|^2  e^{-i \omega_{\mb{k}}t} \mathcal{I}(\mb{k}),
\end{split}
\end{align}
where without loss of generality, hereafter we take this to mean the correlator of the zero-mean variable $\tau \to \tau - \braket{\tau}$, i.e., the connected correlation function. The second line follows from the definition of the vacuum and Eq. \eqref{eq:hmunu}. The function $\mathcal{I}(\mb{k})$ is given in Eq. \eqref{eq:calI}; the main point is that it is $t$-independent, and is rapidly oscillating and decaying like $\sim 1/|\mb{k}|^2$ as $|\mb{k}| \to \infty$.

A common characterization of noise in measurements is the noise power. Using Eq. \eqref{eq:tautau}, we can define the noise power spectral density (PSD) of the $\tau$ observable:
\begin{align}
\label{eq:Stautau-def}
S_{\tau\tau}(\nu) = \int_{-\infty}^{\infty} dt \, e^{i \nu t} \braket{\tau(t) \tau(0)}.
\end{align}
Since the only $t$ dependence in $\braket{\tau(t)\tau(0)}$ is the simple exponential $\sim e^{-i \omega_{\mb{k}}t}$, we can naively exchange the order of integration, do the $dt$ integral, and get a factor $\delta(\nu-\omega_{\mb{k}})$. Doing so, one can perform the resulting momentum integral explicitly and obtain
\be
\label{eq:Stautau-final}
S_{\tau\tau}(\nu) = \frac{\ell_{\rm Pl}^2}{\pi \nu} \Big[ \frac{3 - \cos(2 \nu L)}{6}  - \frac{3 + \cos(2 \nu L)}{2 \nu^2 L^2} + \frac{\sin(2\nu L)}{\nu^3 L^3} \Big],
\ee
with $S_{\tau\tau}(\nu) =0$ for $\nu < 0$. This noise power goes like $\nu$ at low frequencies $\nu \to 0$ and like $1/\nu$ at high frequencies $\nu \to \infty$, so the PSD is finite in both limits. Of course, the most important feature is that it scales like $\ell_{\rm Pl}^2$. 

The noise PSD is directly observable in experiments, and gives a nice interpretation for divergences. If we take a stream of photons and measure $\tau$ continuously, at each time $t$ the outcome is a random variable. With a given set of outcome data $\tau(t)$  for some total integration time $T$, we can form the estimator of the average time-of-flight $\tau_{\rm meas} = \int_0^T dt \, \tau(t)/T$. The variance of this observable is
\be
\braket{ \Delta \tau^2_{\rm meas}} = \frac{2}{\pi}\int_{-\infty}^{\infty} d\nu\, \frac{\sin^2(\nu T/2)}{\nu^2 T^2} S_{\tau\tau}(\nu),
\ee
where we used stationarity of the noise $\braket{\tau(t) \tau(t')} = \braket{\tau(t-t') \tau(0)}$. Using Eq. \eqref{eq:Stautau-final}, this predicts a finite variance for any $T > 0$. The limit $T \to 0$ is logarithmically UV divergent. Here this is because no data was taken.

Technically, to calculate $S_{\tau \tau}(\nu)$, we first need to render $\braket{\tau(t) \tau(0)}$ finite by regulating another high-frequency divergence. The integral \eqref{eq:tautau} diverges logarithmically $\sim \int dk/k$ at large $k$. A straightforward way to regulate the divergence is to take $t \to t - i \epsilon$ in Eq. \eqref{eq:tautau}, which produces a convergent integral. After doing the $dt$ integral, this gives Eq. \eqref{eq:Stautau-final}, except with a factor $S_{\tau\tau} \rightarrow S_{\tau\tau} e^{-\epsilon \nu}$. At this stage, one can either safely take $\epsilon \to 0$, or leave $\epsilon > 0$ finite as a model for an ultraviolet cutoff; even $\epsilon \sim \ell_{\rm Pl}$ does not change the overall $\ell_{\rm Pl}^2$ scaling. This regulator prescription is not arbitrary: it in fact defines \eqref{eq:tautau}, which is a Wightman function (see Appendix \ref{app:geodesics}).

\section{Detector response}
\label{sec:detector}

We now move on to a realistic calculation of the response of an interferometer to the vacuum of quantum gravity. This introduces two new complications. One is that modern interferometric gravitational wave detectors measure distance not by a single round-trip time of light, but with light that reflects many times in a Fabry-Perot cavity. The other is that the detector itself, even in the absence of technical noise, is limited by quantum mechanical noise of its parts, in particular the electromagnetic fluctuations of the light.

\begin{figure}[t]

\begin{tikzpicture}[scale=0.37,photon/.style={decorate,decoration={snake,post length=1mm}}]

\draw [black,fill=lightgray] (-4.5,-2.5) rectangle (-5.5,2.5);
\draw (-5,2.5) -- (-5,3.5);
\draw (-6,4) -- (-4,4);
\draw (-5,3.5) -- (-4.5,4);
\draw (-5,3.5) -- (-5.5,4);

\draw [black,fill=lightgray] (4.5,-2.5) rectangle (5.5,2.5);
\draw (5,2.5) -- (5,3.5);
\draw (4,4) -- (6,4);
\draw (5,3.5) -- (5.5,4);
\draw (5,3.5) -- (4.5,4);

\draw [photon] (-4.5,0) -- (4.5,0);

\node at (-14,1) {laser in};

\draw (-14,0) -- (-5.5,0);
\draw (-12,.5) -- (-11,-.5);
\draw (-11.5,0) -- (-11.5,-3);
\draw (-8,-.5) -- (-7,.5);
\draw (-12,-1.5) -- (-11,-2.5);
\draw (-12.5,-2) -- (-7.5,-2);
\draw (-8,-2.5) -- (-7,-1.5);
\draw (-7.5,0) -- (-7.5,-2);

\draw (-11.8,-1) circle (0.3);
\node at (-13.2,-0.9) {$\tau_{\rm delay}$};

\draw (-12.5,-1.5) -- (-12.5,-2.5);
\draw (-12.5,-1.5) arc(90:270:.5);
\draw (-12,-3) -- (-11,-3);
\draw (-12,-3) arc(180:360:.5);

\node at (-11,-4.5) {measure $\phi(t)$};

\end{tikzpicture}

\caption{Single-arm Fabry-Perot cavity as a gravitational wave detector. The cavity is formed from a partially transparent mirror on the left, and a perfectly reflective mirror on the right, both suspended as harmonic oscillators. The laser light comes into the cavity through the transparent mirror, reflects in the cavity for a time $\sim 1/\kappa$, and then exits. This output light has its phase continuously measured by interfering it with the original laser, forming a homodyne measurement.}
\label{fig:ligo-onearm}
\end{figure}
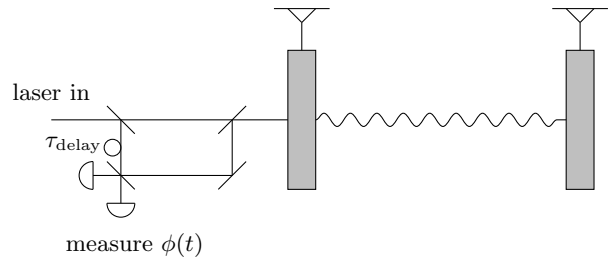

The basic setup is shown in Fig. \ref{fig:ligo-onearm}. We work in a simple setting of a single-arm cavity; this is formally equivalent to the usual two-arm setup used in LIGO and other detectors \cite{Beckey:2023shi}. Laser light of known phase is sent into the cavity. As it transits the cavity its phase changes by an amount proportional to the strain field $h_{\mu\nu}$. After exiting the cavity, this phase is measured by interference with the initial laser field. Here our goal will be to calculate the noise in this phase measurement, including the noise of the gravitational vacuum. The complete Hamiltonian of the system $H = H_{\rm det} + H_{\rm GW} + V_{\rm GW} + H_{\rm I/O}$ describes respectively the detector mirrors and light, the free gravitational perturbations, their coupling to the detector, and the laser and readout system. We give a full accounting of all of these terms in Appendix \ref{app:detector}. 

For our purposes, the main idea is the following: the two mirrors form a Fabry-Perot cavity, and we assume that we measure a single isolated mode with the same frequency $\omega_{\ell}$ as our input laser. We model this mode as
\be
\label{eq:Amu}
A_{\mu}(\mb{x},t) = \frac{\sin(\omega_{\ell} z)}{\sqrt{4 \omega_{\ell} A_{\perp} L}} \left[ \epsilon_{\mu}  e^{-i \omega_{\ell} t} a + \epsilon_{\mu}^* e^{i \omega_{\ell} t} a^\dag \right],
\ee
which is a standing wave with wavevector $k^{\mu} = (\omega_{\ell}, 0, 0, \omega_{\ell})$ and transverse polarization $k_{\mu} \epsilon^{\mu} = 0$. The normalization is so that $a$ is a discrete mode  $[a,a^\dag]=1$, using the cavity length $L$ and a transverse beam area $A_{\perp} = L_{\perp}^2$, the latter of which we will see drop out of the calculation. The cavity electric and magnetic field can be described in terms of its amplitude $X = (a+a^\dag)/\sqrt{2}$ and phase $Y = -i (a- a^\dag)/\sqrt{2}$ quadrature operators:
\begin{align}
\begin{split}
E_x(t,z) & = \sin(\omega_{\ell} z) \sqrt{\frac{\omega_{\ell}}{2 A_{\perp} L}} X(t) \\
B_y(t,z) & = \cos(\omega_{\ell} z) \sqrt{\frac{\omega_{\ell}}{2 A_{\perp} L}} Y(t),
\end{split}
\end{align}
with the specific choice $\epsilon^{\mu} = (0,i,0,0)$, for example. In particular, the phase of the cavity mode means the amount that the mode has rotated to have $Y \sim B_y \neq 0$. In TT gauge, the mirror coordinate positions do not change to $O(h)$ if the mirror is initially at rest, so we can ignore the effect of the gravitational field on the mirror motion to this order~\cite{maggiore2008gravitational,Pang:2018eec}. What we need is the coupling of gravity to the light. Expanding around the coherent laser field $a \to \alpha + a$, the linearized coupling to the electromagnetic fluctuations takes the form
\be
V_{\rm GW} = \frac{1}{2} \int \displaylimits_{\rm cav} d^3\mb{x} \ h_{ij} T^{ij} = F_h X,
\ee
where the ``force'' $F_h$ is given by
\begin{align}
\label{eq:Fh-simple}
F_{h} = \frac{\sqrt{2 \overline{n}} \omega_{\ell} \ell_{\rm Pl}}{V} \int \frac{d^3\mb{k}}{\sqrt{(2\pi)^3 2 E_{\mb{k}}}} \sum_{s} \left[  W_s(\mb{k}) b_{\mb{k},s} +h.c.  \right].
\end{align}
Here $V = A_{\perp} L$ is the volume of the beam, $\overline{n}$ is the average number of laser photons in the cavity, and the window or filtering function
\begin{align}
W_s(\mb{k}) = \frac{\sin(k_x L_{\perp}/2)}{k_x} \frac{\sin(k_y L_{\perp}/2)}{ k_y} \frac{\sin(k_z L/2)}{ k_z} \epsilon_{zz}^s(\mb{k})
\end{align}
encodes the coupling to the gravitational mode ($\mb{k}$,$s$). Here $F_h$ is written as a Schr\"odinger picture operator on the graviton Hilbert space, with units of frequency. This simple expression assumes that the laser $\omega_{\ell} \gg \omega_{\rm GW}$, but does not assume that $\lambda_{\rm GW} \gg L$. In this limit, only the metric perturbations along the beam axis $h_{zz}$ appear [c.f. our geometric result in Eq. \eqref{eq:round-trip}]. See Eq. \eqref{eq:VGW-full} for the complete expression, valid for frequencies above the laser.

The remaining ingredient we need is the input-output fields that couple the readout system to the detector~\cite{gardiner1985input}. The laser and readout are defined by traveling electromagnetic waves entering and exiting the cavity. They are related to the cavity mode through a scattering relation, which in optics is called an I/O relation:
\begin{align}
\begin{split}
\label{eq:IO}
X_{\rm out}(t) & = X_{\rm in}(t) + \sqrt{\kappa} X(t), \\
Y_{\rm out}(t) & = Y_{\rm in}(t) + \sqrt{\kappa} Y(t).
\end{split}
\end{align}
The parameter $\kappa$ has units of a frequency and represents the rate at which photons enter and exit through the transparent mirror. The in/out fields are densities in one dimension and have commutators $[X_{\rm in}(t),X_{\rm in}(t')] = \delta(t-t')$ and so forth. The phase of the light coming out of the cavity is represented by $Y_{\rm out}(t)$. The input fields represent fluctuations around the incoming laser drive, and lead to noise, as discussed below.

Finally, we want to get the observable $Y_{\rm out}$. Using the full Hamiltonian of the system, the Heisenberg equations of motion for the detector work out to a simple linear system (see Appendix \ref{app:detector}). Moving to frequency domain by defining $f(\nu) = \int dt e^{i \nu t} f(t)$ for any operator $f$, we can easily find $Y_{\rm out}(\nu)$. The result is
\begin{align}
\begin{split}
\label{eq:Yout}
Y_{\rm out} & = \sqrt{\kappa} \chi_c F^{h}  \\
& + e^{i \phi_c} Y_{\rm in} + g^2 \kappa \chi_c^2 \chi_m X_{\rm in} + g \sqrt{\kappa} \chi_c \chi_m F_{\rm m,in}.
\end{split}
\end{align}
The cavity and mechanical response functions are
\be
\chi_c(\nu) = \frac{1}{\nu - i \kappa/2}, \ \ \ \chi_m(\nu) = \frac{1}{m [(\nu^2 - \omega_m^2) - i \gamma \nu]}.
\ee
Here $g \sim g_0 \sqrt{\overline{n}}$ and $\gamma$ is the mechanical damping rate of the mirror. The first line of Eq. \eqref{eq:Yout} contains the signal: the  gravitational field ($F_h$) driving the phase of the interferometer light directly. The second line contains the various noise contributions: the shot/phase noise of the input light ($Y_{\rm in}$), the resulting random radiation pressure ($X_{\rm in}$), and finally the thermal fluctuations on the mirror itself ($F_{\rm m,in}$). All of these effects add up to generate the total output light phase $Y_{\rm out}$.

Finally, what we are really interested in is the noise power in the phase of the light coming out of the cavity:
\begin{align}
\begin{split}
\label{eq:SYY-general}
S_{YY}(\nu) & = \int_{-\infty}^{\infty} dt\, e^{i \nu t} \braket{ Y_{\rm out}(t) Y_{\rm out}(0)}_{\rm vac} \\
& = \kappa |\chi_c(\nu)|^2 S_{FF}^{h}(\nu) + {\rm noise~terms}.
\end{split}
\end{align}
Here, the vacuum subscript means that we assumed the same graviton vacuum as above. The power spectrum of the gravitational signal is
\begin{align}
\begin{split}
\label{eq:Shh-def}
S_{FF}^{h}(\nu) & = \int_{-\infty}^{\infty} dt \ e^{i \nu t} \braket{F_h(t) F_h(0)}_{\rm vac} \\
& = \frac{2 \overline{n} \omega_{\ell}^2 \ell_{\rm Pl}^2 \nu}{V^2} \int \frac{d^2\hat{\mb{n}}}{(2\pi)^2} \sum_s |W_s(\nu \hat{\mb{n}})|^2,
\end{split}
\end{align}
where we used Eq. \eqref{eq:Fh-simple} and the free evolution $b_{\mb{k},s} \to e^{-i \omega_{\mb{k}} t} b_{\mb{k},s}$ for the gravitational modes.

\begin{figure}[t]
\includegraphics[width=\linewidth]{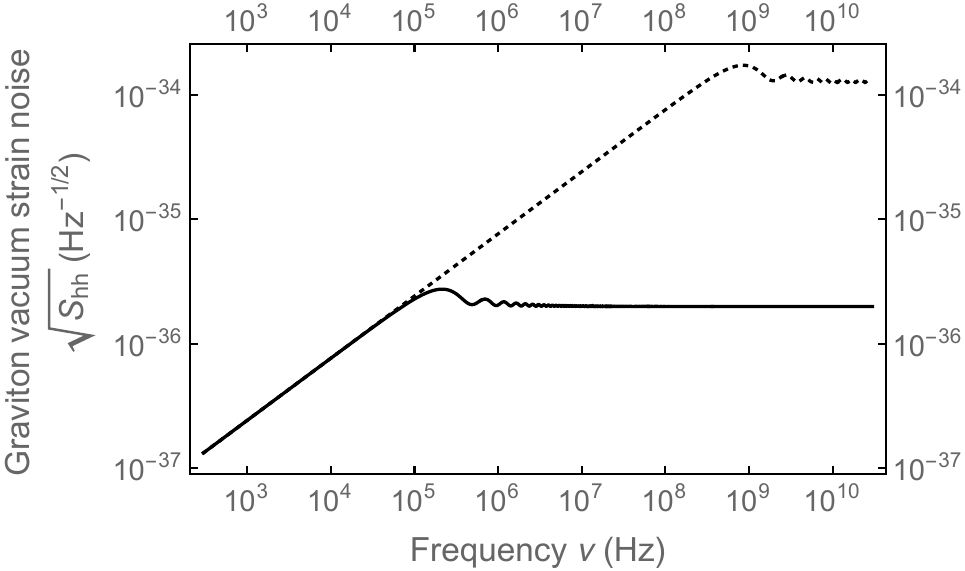}
\caption{Strain noise power of quantum gravitational vacuum fluctuations, expressed in units of strain/$\sqrt{\rm Hz}$. Solid curve: LIGO-like interferometer, with $L \approx 4~{\rm km}$. Dotted: GQuEST-like interferometer, with $L \approx 5~{\rm m}$. The strain PSD is obtained from Eq. \eqref{eq:shh-final} by defining an estimator $h=F_h/\omega_{\ell} \sqrt{\overline{n}} = Y_{\rm out}/\omega_{\ell} \chi_c \sqrt{\kappa \overline{n}}$ [see Eqs. \eqref{eq:hmunu}, \eqref{eq:Fh-simple}, \eqref{eq:Yout}, and \eqref{eq:shh-final}]. For comparison, electromagnetic fluctuations in LIGO's laser give noise around $\sqrt{S_{hh}} \sim 10^{-23}/\sqrt{\rm Hz}$ at 1~{\rm kHz}. The flat behavior at very high frequency is cut off like $1/\nu^5$ for $\nu \gtrsim \omega_{\ell} \approx 300~{\rm THz}$, as discussed in the main text.}
\label{fig:psds}
\end{figure}

Our expression Eq. \eqref{eq:Shh-def} gives the exact noise power spectrum due to the vacuum of perturbative quantum gravity, to leading order in the Planck length. The most important feature, of course, is that it scales like $\ell_{\rm pl}^2$ as advertised. We can analyze some interesting limits of this noise power. Simple Taylor expansion shows that $S_{YY} \sim \nu^1$ as $\nu \to 0$, and $1/\nu^7$ as $\nu \to \infty$, so the noise power is finite in both limits. In a real experiment we will be looking for signals at frequencies $\nu$ such that $\nu \ll 1/L_{\perp}$ and $\nu \ll \omega_{\ell}$ (for example, in LIGO, $L_{\perp} \sim 10~{\rm mm} \sim (30~{\rm GHz})^{-1}$, $\omega_{\ell} \sim 300~{\rm THz}$, and $\nu$ is in the Hz---kHz range \cite{ligo-note}). In this limit, we obtain the simple expression 
\be
\sum_s |W_s(\nu \hat{\mb{n}})|^2 \approx A_{\perp}^2 \frac{\sin^2[\nu L \cos \theta/2]}{\nu^2 \cos^2 \theta} \sin^4\theta.
\ee
Using this we can do the angular integral, giving
\begin{align}
\begin{split}
\label{eq:shh-final}
& S_{YY}(\nu)  \approx \frac{\ell_{\rm Pl}^2}{L^2} \frac{\overline{n} \omega_{\ell}^2 \kappa |\chi_c(\nu)|^2}{\pi \nu} \Big[  \cos (\nu L) - \frac{8}{3} + \frac{\sin(\nu L)}{\nu L} \\
&  - \frac{2 \cos(\nu L)}{(\nu L)^2} + \frac{2 \sin(\nu L)}{(\nu L)^3} + (\nu L) {\rm Si}(\nu L) \Big].
\end{split}
\end{align}
We emphasize that this expression is valid for signals both far below or near and above the free spectral range of the cavity $\nu L \approx \pi/2$ (i.e., frequencies at the lowest optical resonance of the cavity). The former limit is relevant for typical LIGO searches (where $1/L \approx 74~{\rm kHz}$) but the latter is relevant in certain newer, short-baseline devices like GQuEST \cite{vermeulen2021experiment,Vermeulen:2024vgl}, which is looking for MHz-scale signals. In Fig. \ref{fig:psds}, we re-express this as a strain noise for comparison with typical detector sensitivities.

\section{Remarks}

Until direct measurements of quantum gravitational effects are made, a wide range of possible phenomenologies and their predictions for realistic experiments are needed. Our central goal here was to show how a precision calculation in the most conservative theoretical framework---the standard effective field theory of perturbative quantum gravity---can be applied to make predictions in a realistic detector. 

The answer, consistent with basic EFT scaling arguments, is that the effects of quantum gravitational vacuum fluctuations are unobservably small. We give a detailed prediction for the graviton noise power spectral density in Fig. \ref{fig:psds}. In particular, we have studied the divergence structure of the EFT calculation, and shown that there are no UV or IR divergences at this order. This, in particular, implies that that EFT calculation is internally self-consistent, and there is no reason at the level of EFT to expect a large non-perturbative effect. 

Moving forward, we expect that our calculational framework here, especially the detector model, can be used to investigate proposals invoking such large effects, such as~\cite{Hogan:2007hc,hogan2008measurement,Verlinde:2019xfb,Zurek:2020ukz,Li:2022mvy,Bub:2023bfi}, where quantum gravitational effects may be observably large. Our results should also be useful as a starting point to study higher-order (i.e., loop-level) effects and their divergence structure.

\begin{acknowledgments}
We thank Clifford Cheung, Patrick Draper, Temple He, Ted Jacobson, Vincent Lee, Juan Maldacena, Giacomo Marocco, Nick Rodd, Jordan Wilson-Gerow, and Kathryn Zurek for helpful discussions. D.C. and M.K. are supported by the U.S. Department of Energy, Office of High Energy Physics, under Contract No. DEAC02-05CH11231; by DOE Quantum Information Science Enabled Discovery (QuantISED) for High Energy Physics grant KA2401032; and by the Heising-Simons Foundation ``Testing the Quantum Coherence of Gravity'', grant 2023-4467. A. S. is supported by the Heising-Simons Foundation ``Observational Signatures of Quantum Gravity'' collaboration grant 2021-2817; by the D.O.E., Office of High Energy Physics, under Award No. DE-SC0011632; and by the Walter Burke Institute for Theoretical Physics.
\end{acknowledgments}

\bibliography{gravity_vacuum_interferometer}{}

\providecommand{\href}[2]{#2}\begingroup\raggedright\begin{thebibliography}{10}

\bibitem{aasi2015advanced}
J.~Aasi {\em et~al.}, ``Advanced ligo,'' {\em Classical and quantum gravity} {\bfseries 32} no.~7, (2015) 074001.

\bibitem{Aggarwal:2020olq}
N.~Aggarwal {\em et~al.}, ``{Challenges and opportunities of gravitational-wave searches at MHz to GHz frequencies},'' \href{http://dx.doi.org/10.1007/s41114-021-00032-5}{{\em Living Rev. Rel.} {\bfseries 24} no.~1, (2021) 4}, \href{http://arxiv.org/abs/2011.12414}{{\ttfamily arXiv:2011.12414 [gr-qc]}}.

\bibitem{Aggarwal:2020umq}
N.~Aggarwal, G.~P. Winstone, M.~Teo, M.~Baryakhtar, S.~L. Larson, V.~Kalogera, and A.~A. Geraci, ``{Searching for New Physics with a Levitated-Sensor-Based Gravitational-Wave Detector},'' \href{http://dx.doi.org/10.1103/PhysRevLett.128.111101}{{\em Phys. Rev. Lett.} {\bfseries 128} no.~11, (2022) 111101}, \href{http://arxiv.org/abs/2010.13157}{{\ttfamily arXiv:2010.13157 [gr-qc]}}.

\bibitem{vermeulen2021experiment}
S.~M. Vermeulen, L.~Aiello, A.~Ejlli, W.~L. Griffiths, A.~L. James, K.~L. Dooley, and H.~Grote, ``An experiment for observing quantum gravity phenomena using twin table-top 3d interferometers,'' {\em Classical and Quantum Gravity} {\bfseries 38} no.~8, (2021) 085008.

\bibitem{Vermeulen:2024vgl}
S.~M. Vermeulen {\em et~al.}, ``{Photon Counting Interferometry to Detect Geontropic Space-Time Fluctuations with GQuEST},'' \href{http://arxiv.org/abs/2404.07524}{{\ttfamily arXiv:2404.07524 [gr-qc]}}.

\bibitem{Hogan:2007hc}
C.~J. Hogan, ``{Spacetime Indeterminacy and Holographic Noise},'' \href{http://arxiv.org/abs/0706.1999}{{\ttfamily arXiv:0706.1999 [gr-qc]}}.

\bibitem{hogan2008measurement}
C.~J. Hogan, ``Measurement of quantum fluctuations in geometry,'' {\em Physical Review D} {\bfseries 77} no.~10, (2008) 104031.

\bibitem{Verlinde:2019xfb}
E.~P. Verlinde and K.~M. Zurek, ``{Observational signatures of quantum gravity in interferometers},'' \href{http://dx.doi.org/10.1016/j.physletb.2021.136663}{{\em Phys. Lett. B} {\bfseries 822} (2021) 136663}, \href{http://arxiv.org/abs/1902.08207}{{\ttfamily arXiv:1902.08207 [gr-qc]}}.

\bibitem{Verlinde:2019ade}
E.~Verlinde and K.~M. Zurek, ``{Spacetime Fluctuations in AdS/CFT},'' \href{http://dx.doi.org/10.1007/JHEP04(2020)209}{{\em JHEP} {\bfseries 04} (2020) 209}, \href{http://arxiv.org/abs/1911.02018}{{\ttfamily arXiv:1911.02018 [hep-th]}}.

\bibitem{Zurek:2020ukz}
K.~M. Zurek, ``{On vacuum fluctuations in quantum gravity and interferometer arm fluctuations},'' \href{http://dx.doi.org/10.1016/j.physletb.2022.136910}{{\em Phys. Lett. B} {\bfseries 826} (2022) 136910}, \href{http://arxiv.org/abs/2012.05870}{{\ttfamily arXiv:2012.05870 [hep-th]}}.

\bibitem{Li:2022mvy}
D.~Li, V.~S.~H. Lee, Y.~Chen, and K.~M. Zurek, ``{Interferometer response to geontropic fluctuations},'' \href{http://dx.doi.org/10.1103/PhysRevD.107.024002}{{\em Phys. Rev. D} {\bfseries 107} no.~2, (2023) 024002}, \href{http://arxiv.org/abs/2209.07543}{{\ttfamily arXiv:2209.07543 [gr-qc]}}.

\bibitem{Bub:2023bfi}
M.~W. Bub, Y.~Chen, Y.~Du, D.~Li, Y.~Zhang, and K.~M. Zurek, ``{Quantum gravity background in next-generation gravitational wave detectors},'' \href{http://dx.doi.org/10.1103/PhysRevD.108.064038}{{\em Phys. Rev. D} {\bfseries 108} no.~6, (2023) 064038}, \href{http://arxiv.org/abs/2305.11224}{{\ttfamily arXiv:2305.11224 [gr-qc]}}.

\bibitem{Donoghue:1995cz}
J.~F. Donoghue, ``{Introduction to the effective field theory description of gravity},'' in {\em {Advanced School on Effective Theories}}.
\newblock 6, 1995.
\newblock \href{http://arxiv.org/abs/gr-qc/9512024}{{\ttfamily arXiv:gr-qc/9512024}}.

\bibitem{Burgess:2003jk}
C.~P. Burgess, ``{Quantum gravity in everyday life: General relativity as an effective field theory},'' \href{http://dx.doi.org/10.12942/lrr-2004-5}{{\em Living Rev. Rel.} {\bfseries 7} (2004) 5--56}, \href{http://arxiv.org/abs/gr-qc/0311082}{{\ttfamily arXiv:gr-qc/0311082}}.

\bibitem{Donoghue:2022eay}
J.~F. Donoghue, {\em {Quantum General Relativity and Effective Field Theory}}.
\newblock 2023.
\newblock \href{http://arxiv.org/abs/2211.09902}{{\ttfamily arXiv:2211.09902 [hep-th]}}.

\bibitem{Jaekel:1993ft}
M.-T. Jaekel and S.~Reynaud, ``{Gravitational quantum limit for length measurements},'' \href{http://dx.doi.org/10.1016/0375-9601(94)90838-9}{{\em Phys. Lett. A} {\bfseries 185} (1994) 143--148}, \href{http://arxiv.org/abs/quant-ph/9801074}{{\ttfamily arXiv:quant-ph/9801074}}.

\bibitem{Pang:2018eec}
B.~Pang and Y.~Chen, ``{Quantum interactions between a laser interferometer and gravitational waves},'' \href{http://dx.doi.org/10.1103/PhysRevD.98.124006}{{\em Phys. Rev. D} {\bfseries 98} no.~12, (2018) 124006}, \href{http://arxiv.org/abs/1808.09122}{{\ttfamily arXiv:1808.09122 [quant-ph]}}.

\bibitem{Parikh:2020nrd}
M.~Parikh, F.~Wilczek, and G.~Zahariade, ``{The Noise of Gravitons},'' \href{http://dx.doi.org/10.1142/S0218271820420018}{{\em Int. J. Mod. Phys. D} {\bfseries 29} no.~14, (2020) 2042001}, \href{http://arxiv.org/abs/2005.07211}{{\ttfamily arXiv:2005.07211 [hep-th]}}.

\bibitem{Parikh:2020kfh}
M.~Parikh, F.~Wilczek, and G.~Zahariade, ``{Quantum Mechanics of Gravitational Waves},'' \href{http://dx.doi.org/10.1103/PhysRevLett.127.081602}{{\em Phys. Rev. Lett.} {\bfseries 127} no.~8, (2021) 081602}, \href{http://arxiv.org/abs/2010.08205}{{\ttfamily arXiv:2010.08205 [hep-th]}}.

\bibitem{Parikh:2020fhy}
M.~Parikh, F.~Wilczek, and G.~Zahariade, ``{Signatures of the quantization of gravity at gravitational wave detectors},'' \href{http://dx.doi.org/10.1103/PhysRevD.104.046021}{{\em Phys. Rev. D} {\bfseries 104} no.~4, (2021) 046021}, \href{http://arxiv.org/abs/2010.08208}{{\ttfamily arXiv:2010.08208 [hep-th]}}.

\bibitem{Hertzberg:2021rbl}
M.~P. Hertzberg and J.~A. Litterer, ``{Bound on quantum fluctuations in gravitational waves from LIGO-Virgo},'' \href{http://dx.doi.org/10.1088/1475-7516/2023/03/009}{{\em JCAP} {\bfseries 03} (2023) 009}, \href{http://arxiv.org/abs/2112.12159}{{\ttfamily arXiv:2112.12159 [gr-qc]}}.

\bibitem{Guerreiro:2021qgk}
T.~Guerreiro, F.~Coradeschi, A.~M. Frassino, J.~R. West, and E.~Schioppa, Junior., ``{Quantum signatures in nonlinear gravitational waves},'' \href{http://dx.doi.org/10.22331/q-2022-12-19-879}{{\em Quantum} {\bfseries 6} (2022) 879}, \href{http://arxiv.org/abs/2111.01779}{{\ttfamily arXiv:2111.01779 [gr-qc]}}.

\bibitem{Carney:2023nzz}
D.~Carney, V.~Domcke, and N.~L. Rodd, ``{Graviton detection and the quantization of gravity},'' \href{http://dx.doi.org/10.1103/PhysRevD.109.044009}{{\em Phys. Rev. D} {\bfseries 109} no.~4, (2024) 044009}, \href{http://arxiv.org/abs/2308.12988}{{\ttfamily arXiv:2308.12988 [hep-th]}}.

\bibitem{weinberg1964photons}
S.~Weinberg, ``Photons and gravitons in s-matrix theory: derivation of charge conservation and equality of gravitational and inertial mass,'' {\em Physical Review} {\bfseries 135} no.~4B, (1964) B1049.

\bibitem{weinberg1965photons}
S.~Weinberg, ``Photons and gravitons in perturbation theory: Derivation of maxwell's and einstein's equations,'' {\em Physical Review} {\bfseries 138} no.~4B, (1965) B988.

\bibitem{Maldacena:1997re}
J.~M. Maldacena, ``{The Large N limit of superconformal field theories and supergravity},'' \href{http://dx.doi.org/10.4310/ATMP.1998.v2.n2.a1}{{\em Adv. Theor. Math. Phys.} {\bfseries 2} (1998) 231--252}, \href{http://arxiv.org/abs/hep-th/9711200}{{\ttfamily arXiv:hep-th/9711200}}.

\bibitem{Heemskerk:2009pn}
I.~Heemskerk, J.~Penedones, J.~Polchinski, and J.~Sully, ``{Holography from Conformal Field Theory},'' \href{http://dx.doi.org/10.1088/1126-6708/2009/10/079}{{\em JHEP} {\bfseries 10} (2009) 079}, \href{http://arxiv.org/abs/0907.0151}{{\ttfamily arXiv:0907.0151 [hep-th]}}.

\bibitem{Dimopoulos:2007cj}
S.~Dimopoulos, P.~W. Graham, J.~M. Hogan, M.~A. Kasevich, and S.~Rajendran, ``{Gravitational Wave Detection with Atom Interferometry},'' \href{http://dx.doi.org/10.1016/j.physletb.2009.06.011}{{\em Phys. Lett. B} {\bfseries 678} (2009) 37--40}, \href{http://arxiv.org/abs/0712.1250}{{\ttfamily arXiv:0712.1250 [gr-qc]}}.

\bibitem{Lee:2024oxo}
V.~S.~H. Lee and K.~M. Zurek, ``{Proper Time Observables of General Gravitational Perturbations in Laser Interferometry-based Gravitational Wave Detectors},'' \href{http://arxiv.org/abs/2408.03363}{{\ttfamily arXiv:2408.03363 [hep-ph]}}.

\bibitem{Sivaramakrishnan:2024ydy}
A.~Sivaramakrishnan, ``{Correlators of Worldline Proper Length},'' \href{http://arxiv.org/abs/2406.17205}{{\ttfamily arXiv:2406.17205 [hep-th]}}.

\bibitem{Maggiore:2007ulw}
M.~Maggiore, \href{http://dx.doi.org/10.1093/acprof:oso/9780198570745.001.0001}{{\em {Gravitational Waves. Vol. 1: Theory and Experiments}}}.
\newblock Oxford University Press, 2007.

\bibitem{Beckey:2023shi}
J.~Beckey, D.~Carney, and G.~Marocco, ``{Quantum measurements in fundamental physics: a user's manual},'' \href{http://arxiv.org/abs/2311.07270}{{\ttfamily arXiv:2311.07270 [hep-ph]}}.

\bibitem{maggiore2008gravitational}
M.~Maggiore, {\em Gravitational waves}, vol.~2.
\newblock Oxford university press, 2008.

\bibitem{gardiner1985input}
C.~W. Gardiner and M.~J. Collett, ``Input and output in damped quantum systems: Quantum stochastic differential equations and the master equation,'' {\em Physical Review A} {\bfseries 31} no.~6, (1985) 3761.

\bibitem{ligo-note}
``Ligo technical document ligo-t0900043.''
\newblock \url{https://dcc.ligo.org/public/0000/T0900043/011/LIGO-T0900043-11.pdf}.

\bibitem{Hartman:2015lfa}
T.~Hartman, S.~Jain, and S.~Kundu, ``{Causality Constraints in Conformal Field Theory},'' \href{http://dx.doi.org/10.1007/JHEP05(2016)099}{{\em JHEP} {\bfseries 05} (2016) 099}, \href{http://arxiv.org/abs/1509.00014}{{\ttfamily arXiv:1509.00014 [hep-th]}}.

\bibitem{Birrell_Davies_1982}
N.~D. Birrell and P.~C.~W. Davies, {\em Quantum Fields in Curved Space}.
\newblock Cambridge Monographs on Mathematical Physics. Cambridge University Press, 1982.

\bibitem{Kravchuk:2018htv}
P.~Kravchuk and D.~Simmons-Duffin, ``{Light-ray operators in conformal field theory},'' \href{http://dx.doi.org/10.1007/JHEP11(2018)102}{{\em JHEP} {\bfseries 11} (2018) 102}, \href{http://arxiv.org/abs/1805.00098}{{\ttfamily arXiv:1805.00098 [hep-th]}}.

\bibitem{clerk2010introduction}
A.~A. Clerk, M.~H. Devoret, S.~M. Girvin, F.~Marquardt, and R.~J. Schoelkopf, ``Introduction to quantum noise, measurement, and amplification,'' {\em Reviews of Modern Physics} {\bfseries 82} no.~2, (2010) 1155--1208.

\bibitem{whittle2023unification}
C.~Whittle, L.~McCuller, V.~Sudhir, and M.~Evans, ``Unification of thermal and quantum noises in gravitational-wave detectors,'' {\em Physical Review Letters} {\bfseries 130} no.~24, (2023) 241401.

\end{thebibliography}\endgroup
\bibliographystyle{utphys}

\clearpage

\appendix

\section{Geodesics and singularities}
\label{app:geodesics}

Here we give a few more details in the geometric calculations of Sec. I, and in particular some discussion on the choice of $i \epsilon$ prescription. 

The explicit  flat-space null geodesic path for a photon emitted at $t$ used in the main text is given by the piecewise function on $0 \leq \eta \leq 2L$:
\be
\gamma_t^{\mu}(\eta) = \begin{pmatrix} t + \eta \\ z_L + \eta \end{pmatrix}, \ \begin{pmatrix} t  + \eta \\ z_L + 2L - \eta \end{pmatrix}
\ee
for $0 \leq \eta \leq L$ and $L \leq \eta \leq 2L$, respectively. In the presence of metric fluctuations the null geodesic path receives $O(h)$ corrections (see Fig. \ref{fig:interferometer}), even in the transverse $x,y$-directions. Straightforward power counting shows that for proper time $\tau(t_{1})$, the transverse motion contributes to $O(h^2)$ and only the flat-space path parametrization is required on the right-hand side of Eq. \eqref{eq:round-trip}. Rewriting the integrals in the first line of Eq. \eqref{eq:tautau} with this affine parametrization and performing the affine integrals, one obtains the second line of Eq. \eqref{eq:tautau}, with
\begin{align}
\begin{split}
\label{eq:calI}
\mathcal{I}(\mb{k}) & = \Big[ \frac{\sin^2[(\omega_{\mb{k}} - k_z)L/2]}{(\omega_{\mb{k}} - k_z)^2} + \frac{\sin^2[(\omega_{\mb{k}} + k_z)L/2]}{(\omega_{\mb{k}} + k_z)^2} \\
& + \frac{(\cos k_z L - \cos \omega_{\mb{k}} L) \cos \omega_{\mb{k}} L}{\omega_{\mb{k}}^2 - k_z^2} \Big].
\end{split}
\end{align}
The first line here comes from the integration regions with both paths moving left or both right, while the second line comes from the two counter-propagating integration regions. The polarization sum $\sum_{s}|\epsilon_{zz}^s(\mb{k})|^2 = \sin^4(\theta)$ is standard~\cite{maggiore2008gravitational}. Using this, $k_z = \omega_{\mb{k}} \cos \theta$, and Eq. \eqref{eq:calI}, we see that the polar integral in the noise PSD is trivial. The $d\theta$ is not, but it can be performed without too much trouble, and in particular there is no singularity at $\cos \theta = 1$. The result is Eq. \eqref{eq:Stautau-final}.

The $\braket{\tau(t)\tau(0)}$ correlator \eqref{eq:tautau}
is a Wightman function, and using its precise definition helps perform integrals that are otherwise naively divergent. This is analogous to how the integral $\delta(x)=\frac{1}{2\pi}\int_{-\infty}^{-\infty} dk~ e^{ikx}$ cannot be directly evaluated, but the regulated version can be: $\delta_\epsilon(x)=\frac{1}{2\pi}\int_{-\infty}^\infty dk e^{ikx-\epsilon |k|}= \frac{1}{\pi}\frac{ \epsilon}{x^2+\epsilon^2}$. Like the Dirac delta function, Wightman functions are distributions, and an $i\epsilon$ prescription defines Wightman functions in Lorentzian signature as the boundary value of analytically-continued Euclidean Wightman functions. This $i\epsilon$ prescription rigorously regulates the large-$k$ divergence arising in the integral over $\mathcal{I}(\mb{k})$, which makes the Fourier transform in \eqref{eq:Stautau-def} subtlety-free.

To illustrate, consider a free massless scalar field theory in four spacetime dimensions. With $x = (x^0,\mb{x})$, the two-point vacuum Wightman function $\braket{\phi(x) \phi(0)}$ is defined as $\braket{\phi(x) \phi(0)} \equiv \lim_{\epsilon \rightarrow 0} \braket{\phi(x^0-i \epsilon, \mb{x}) \phi(0)}$, and
\begin{equation}
\braket{\phi(x^0-i \epsilon, \mb{x}) \phi(0)}  = \int d^4 p \frac{e^{i (p^0 (i x^0 + \epsilon) + \mb{p} \cdot \mb{x}) }}{(p^0)^2+\mb{p}^2},
\end{equation}
where $p = (p^0,\mb{p})$ is the Euclidean momentum with $p^0$ integrated between $\pm \infty$ and Euclidean time $\tau$ in $e^{i p^0 \tau}$ has been analytically continued to $\tau \rightarrow i (x^0 - i\epsilon)$, where $x^0$ is the desired Lorentzian time. We neglect overall numerical factors for brevity. The small Euclidean time $\epsilon$ allows us to close the $p^0$ contour above and pick up the $p^0 = i |\mb{p}| $ pole, giving
\begin{equation}
\braket{\phi(x^0-i \epsilon, \mb{x})\phi(0)} = \int \frac{d^3 p}{|\mb{p}|}
e^{i (|\mb{p}| (- x^0  +i\epsilon)+  \mb{p} \cdot \mb{x})} ,
\end{equation}
an integral over positive-energy momenta. Doing the angular integrals and changing variables to $k =|\mb{p}|$,
\begin{align}
\braket{\phi(x^0-i \epsilon, \mb{x})\phi(0)} =& \frac{1}{|\mb{x}|} \int_0^\infty d k \sin(k|\mb{x}|) e^{-i k  x^0 - k \epsilon} 
\nonumber
\\
=& \frac{1}{\mb{x}^2 - (x^0 - i \epsilon)^2},
\end{align}
the expected position-space Wightman function. We see that keeping $\epsilon$ non-zero was crucial for rendering the above integral well-defined. As long as $x^2 \neq 0$, we can now safely set $\epsilon = 0$. For further review, see \cite{Hartman:2015lfa, Birrell_Davies_1982}.

Applying this prescription to the proper-time two-point function yields the well-defined quantity 
\begin{equation}
\braket{\tau(t)\tau(0)} \equiv \lim_{\epsilon \rightarrow 0^+} \braket{ \tau(t-i\epsilon)\tau(0)}.
\end{equation}
With $\epsilon \neq 0$, $\braket{ \tau(t-i\epsilon)\tau(0)}$ can be computed by performing the momentum integral, and it has logarithmic singularities in $t$ regulated by $\epsilon$. See \cite{Kravchuk:2018htv} for detailed analysis of light-ray transforms of Wightman functions.

\section{Detector details}
\label{app:detector}

In this appendix we give a more detailed account of the detector model. The treatment follows standard references on quantum optomechanics, gravitational waves, and input-output theory; see for example~\cite{clerk2010introduction,Beckey:2023shi} and references therein. 

We begin by deriving the complete Hamiltonian $H = H_{\rm det} + V_{\rm GW} + H_{\rm GW} + H_{\rm I/O}$. The detector Hamiltonian describes the mirrors, which are suspended as harmonic oscillators of very low frequency $\omega_m$, and the light between them, which forms a discrete set of optical modes with frequencies $\omega_n = n \pi/L_{\rm cav}(z)$. Here $L_{\rm cav}(z) \approx L + z$ is the length of the cavity in terms of the relative displacement $z \ll L$ of the mirrors from their equilibrium. This generates an opto-mechanical coupling between the optical field and the mirror motion. Together, we have
\be
\label{eq:H-det}
H_{\rm det} = (\omega_{\ell} + g_0 z) a^\dag a + \frac{p^2}{2m} + \frac{1}{2} m \omega_m^2 z^2.
\ee
We are focusing on just the mode at the laser frequency, and assumed that $\omega_{\ell} = \omega_n(0)$ for some $n \gg 1$, i.e., the laser is on resonance with some particular equilibrium cavity mode. In this case, $g_0 = \omega_{\ell}/L$ is the optomechanical coupling strength.

Next we discuss the gravitational terms. The kinetic term
\be
H_{\rm GW} = \frac{1}{2\ell_{\rm Pl}^2}\int d^3\mb{x} \,[\dot h_{\mu\nu}\dot h^{\mu\nu}+ \sum_{i}\partial_{i}h_{\mu\nu}\partial_{i}h^{\mu\nu}]+ \dots
\ee
has the usual plane wave eigenstates. The interaction term is
\be
V_{\rm GW} = \frac{1}{2} \int \displaylimits_{\rm cavity} d^3\mb{x}\, h_{\mu\nu} T^{\mu\nu}.
\ee
As discussed in the main text, in TT gauge, since our mirrors are taken to be in their ground state and thus approximately at rest (the uncertainty is $\Delta v \sim \sqrt{\omega_m/m} \sim 10^{-27}$ with $m = 40~{\rm kg}$ and $\omega_m = 1~{\rm Hz}$), we only need to worry about the coupling to the stress tensor of the optical field. Focusing on the cavity mode of interest, defined in Eq. \eqref{eq:Amu}, the only non-zero stress-energy tensor components we need are
\begin{align}
\begin{split}
\label{eq:Tzz}
T^{xx} & = T_0 \left[ 2 \{ a,a^\dag \} \cos 2 \omega_{\ell} z - a^2 - a^{\dag 2}  \right] = - T^{yy} \\
T^{zz} & = T_0 \Big[ 2 \{ a,a^\dag \} - a^2  \cos 2 \omega_{\ell} z - a^{\dag 2}  \cos 2 \omega_{\ell} z  \Big].
\end{split}
\end{align}
Here $T_0 = \omega_{\ell}/8 A_{\perp} L$, the off-diagonal spatial components are zero, and we have written this in the Schr\"odinger picture by setting all the $e^{\pm i \omega_{\ell} t}$ factors to $1$ after taking time derivatives. Inserting Eqs. \eqref{eq:Tzz} and \eqref{eq:hmunu} into $V_{\rm GW}$, we obtain
\begin{align}
\begin{split}
\label{eq:VGW-full}
& V_{\rm GW} = \frac{T_0 \ell_{\rm Pl}}{2} \int \displaylimits_{\rm cav} d^3\mb{x} \int \frac{d^3\mb{k}}{\sqrt{(2\pi)^3 2 E_{\mb{k}}}} \sum_{s} \\
& \times \Big[ \left( \epsilon^s_{xx}(\mb{k}) - \epsilon^s_{yy}(\mb{k}) \right) \left(2 \{ a,a^\dag \} \cos 2 \omega_{\ell} z -a^2 - a^{\dag 2}  \right) \\
& + \epsilon^s_{zz}(\mb{k}) \left(2 \{ a,a^\dag \}- a^2  \cos 2 \omega_{\ell} z  -  a^{\dag 2}  \cos 2 \omega_{\ell} z  \right) \Big] \\
& \times e^{i \mb{k} \cdot \mb{x}} b_{\mb{k},s} + h.c.
\end{split}
\end{align}
The spatial integrals can be done easily:
\begin{align}
\begin{split}
& \int d^2\mb{x}_{\perp} \, e^{i \mb{k}_{\perp} \cdot \mb{x}_{\perp}} = 4 \frac{\sin k_x L_{\perp}/2}{k_x} \frac{\sin k_y L_{\perp}/2}{k_y} \\
& \int_0^L dz \, e^{i k_z z}  = 2\frac{ \sin k_z L/2}{k_z}  \\
& \int_0^L dz \, e^{i k_z z} \cos 2 \omega_{\ell} z = 2 \frac{k_z^2}{4 \omega_{\ell}^2 - k_z^2} \frac{ \sin k_z L/2}{k_z}.
\end{split}
\end{align}
We have dropped an overall phase in these expressions which cancels out of $|W|^2$.

At this stage, to simplify things, we now make the approximation that the laser frequency $\omega_{\ell}$ is the fastest frequency of interest, in particular $\omega_{\ell} \gg \nu$ where $\nu$ is in the detection band. This allows us to drop the terms with $\cos 2 \omega_{\ell} z$ dependence. It also allows us to make the ``rotating wave approximation'', common in quantum optics: terms of the form $a^2$ and $a^{\dag 2}$ are dropped, because these rotate very fast compared to the time-independent $a^\dag a$ type terms, and therefore average out over an observation period~\cite{clerk2010introduction}. Put together, this leaves us with
\begin{align}
V_{\rm GW} = 8 T_0 \ell_{\rm Pl} \int \frac{d^3\mb{k}}{\sqrt{(2\pi)^3 2 E_{\mb{k}}}} \sum_{s} W_s(\mb{k}) b_{\mb{k},s} \{ a,a^\dag \}  + h.c.
\end{align}
This expression is valid for any state. When we include the laser drive, the cavity mode is displaced to $a \to \alpha + a$, where $|\alpha|^2 = \overline{n} = P_{\rm in}/\omega_{\ell} \kappa$ is the number of photons circulating in the cavity in terms of the input laser power $P_{\rm in}$ and cavity loss rate $\kappa$. Expanding around this drive $|\alpha| \gg 1$ we now keep the terms linear in $a$. Taking $\alpha$ to be real by appropriately choosing the laser phase, the result is
\be
V_{\rm GW} = F_h X
\ee
with $F_h$ given in Eq. \eqref{eq:Fh-simple}.

Finally, we need to discuss the input/output fields $H_{\rm I/O}$. In brief, these form a continuum of modes that couple to the cavity field, namely the electromagnetic modes along the laser beam axis outside the cavity. One can follow these microscopically using a scattering treatment \cite{gardiner1985input}, where the late-time Heisenberg fields are related to the earlier-time ones through Eq. \eqref{eq:IO}. Following the standard treatment of these I/O fields and the rest of the system, one can now work out the Heisenberg equations of motion for all the relevant system operators. The result is (see, e.g., \cite{Beckey:2023shi} for a step-by-step derivation):
\begin{align}
\begin{split}
\label{eq:EOM}
\dot{X} & = -\frac{\kappa}{2} X + \sqrt{\kappa} X_{\rm in} \\
\dot{Y} & = -\frac{\kappa}{2} Y + \sqrt{\kappa} Y_{\rm in} + F_{h} + g x \\
\dot{x} & = \frac{p}{m} \\
\dot{p} & = -m \omega_m^2 x - \gamma p + F_{\rm m,in} + g X.
\end{split}
\end{align}
Here $g = g_0 \sqrt{\overline{n}}/z_0$, the effective coupling strength of the mirror-light interaction in the presence of the laser drive, where $z_0 = 1/\sqrt{2 m \omega_m}$ is the mirror's ground-state uncertainty. We also included mechanical damping $\gamma \ll \omega_m$ on the mirror, which is driven by the random force $F_{\rm m,in}$, which represents the effect of thermal forces on the mirror, including vacuum fluctuations when $T = 0$ \cite{whittle2023unification}. The solution to these equations for $Y_{\rm out}$ is given in Eq. \eqref{eq:Yout}.

In Eq. \eqref{eq:SYY-general}, we only explicitly reported the part of the output phase coming from the graviton vacuum noise, $S_{FF}^h$. Most of the rich physics of quantum-limited interferometers is in the other terms in Eq. \eqref{eq:Yout}, i.e., the noise coming from the detector itself. To reproduce the usual noise curves in these detectors, essentially one just adds the terms in Eq. \eqref{eq:Yout} in quadrature. Assuming that the laser fluctuations $Y_{\rm in}, X_{\rm in}$ are in their quantum vacuum, the result is a good approximation of the real noise curves in a quantum-limited interferometer. Again, see \cite{Beckey:2023shi} for a detailed treatment.

\end{document}